\documentclass[preprint,showpacs,preprintnumbers,amsmath,amssymb,dvips]{revtex4}

\usepackage{graphicx}% Include figure files
\usepackage{dcolumn}% Align table columns on decimal point
\usepackage{bm}% bold math

\begin{document}

\preprint{DISTA-FIN-08-01}

\title{A random telegraph signal of Mittag-Leffler type}% Force line breaks with \\

\author{Simone Ferraro}
\email{sf349@cam.ac.uk}
 \affiliation{Queens' College, University of Cambridge, Cambridge, UK}%Lines break automatically or can be forced with \\
 \author{Michele Manzini}
 \email{michele.manzini@mfn.unipmn.it}
\author{Aldo Masoero}
 \email{aldo.masoero@mfn.unipmn.it}
\author{Enrico Scalas}
 \email{enrico.scalas@mfn.unipmn.it}
\affiliation{Dipartimento di Scienze e Tecnologie Avanzate, Universit\`a del Piemonte Orientale, Alessandria, Italy}

%Authors' institution and/or address\\
%This line break forced with \textbackslash\textbackslash
%}%

%\author{Charlie Author}
 %\homepage{http://www.Second.institution.edu/~Charlie.Author}
%\affiliation{
%Second institution and/or address\\
%This line break forced% with \\
%}%

\date{\today}% It is always \today, today,
             %  but any date may be explicitly specified

\begin{abstract}
A general method is presented to explicitly compute autocovariance functions for
non-Poisson dichotomous noise based on renewal theory. The method is specialized
to a random telegraph signal of Mittag-Leffler type. Analytical predictions are
compared to Monte Carlo simulations. Non-Poisson dichotomous noise is
non-stationary and standard spectral methods fail to describe it properly as
they assume stationarity. 
\end{abstract}

% PACS, the Physics and Astronomy
                             % Classification Scheme.
%\keywords{Random telegraph signal, dichotomous noise, Mittag-Leffler function}%Use showkeys class option if keyword
                              %display desired
\maketitle

\section{Introduction}

%{\em Put some generalities here}
A complete theory of stochastic processes was formulated by A.N. Kolmogorov in 1933 \cite{kolmogorov33}.
It included the characterization of stochastic processes in terms of finite-dimensional distribution
functions \cite{kolmogorovnote}. 

Since the early XXth century, it had been clear that the theory of
stochastic processes has important applications in electronics and communications, for the basic
understanding of devices such as vacuum tubes and, later, solid state diodes and transistors
as well as for the behavior of cable and wireless communication systems.
In particular, it turned out that noise can be described in terms by equations like the following one
\begin{equation}
\label{shotnoise}
I(t) = \sum_{k=-\infty}^{+\infty} F(t - t_k)
\end{equation}
where $F(t)$ is the current produced by an electron reaching the anode of a vacuum tube
at time $t$, the $k$-th electron arrives at a random time $t_k$
and the series (\ref{shotnoise}) is supposed to converge. S.O. Rice gave
an account of the early developments in a Bell Labs monograph
published in 1944 \cite{rice44}.

Rice considers the example of the so-called random telegraph noise (RTN) or random telegraph signal (RTS),
a stationary dichotomous noise where a current or a voltage randomly flips between two levels
$\pm a$ as shown in Fig. 1 and flip events
follow a Poisson process. An early treatment of random telegraph signals can be found in Kenrick's paper
of 1929 \cite{kenrick29}. According to Rice, Kenrick was one of the first authors
to use correlogram methods to compute the power spectrum of a signal $X(t)$.
This was done by first obtaining the autocorrelation function
\begin{equation}
\label{introautocor} 
R_{XX} (\Delta t;t) = \mathbb{E} [(X(t) - \mathbb{E} [X(t)]) (X(t+\Delta t) - \mathbb{E} [X(t+\Delta t)])],
\end{equation}
where $\mathbb{E}[X]$ denotes the expected value of the random variable $X$.
Based on results by Wiener, Khintchin and Cram\'er, it is possible to show that, for a stationary
signal when $R_{XX}$ only depends on the lag $\Delta t$, the power spectrum is the Fourier transform of the autocorrelation function
\cite{khintchine34,cramer40}:
\begin{equation}
\label{wienerkhintchine}
\hat{S}_{XX} (\omega) = \int_{-\infty}^{+\infty} \exp(i \omega v) R_{XX} (v) dv.
\end{equation}
For RTS with $a=1$ and unitary average waiting time (a.k.a. duration), one gets
\begin{equation}
\label{introRTSautocor}
R_{XX} = \exp(-2 |\Delta t|)
\end{equation}
a results which will be derived below for positive $\Delta t$, leading to a Lorentzian power spectrum
\begin{equation}
\label{lorentzian}
\hat{S} (\omega) = \frac{4}{4+\omega^2}.
\end{equation}

Even if very simple, RTS has interesting applications in different fields. 
A complete survey of the literature on RTS and its applications is beyond the
scope of the present paper. However, it is interesting to remark that RTS has
been used to explain 1/$f$ noise in electronic devices. Indeed, A. Mc Whorter
showed that an appropriate superposition of Lorentzian spectra for RTSs of different
average duration gives 1/$f$ noise \cite{mcwhorter55,stepanescu74,milotti}.

More recently, P. Allegrini {\em et al.} studied non-Poisson dichotomous noise 
\cite{allegrini04}. Their paper is particularly relevant here, as 
a particular case of non-Poisson dichotomous noise is studied below in some detail, namely
a generalized RTS where waiting times between consecutive level changes follow
a Mittag-Leffler distribution. This particular process never reaches
a stationary state in finite time and, for this reason, standard spectral
methods fail to properly represent its properties.

Our paper is organized as follows. In section II, devoted to theory,
the stationary case is briefly reviewed. Then some tools are introduced
taken from renewal theory. These methods lead to a general formula
for the autocovariance function of a generalized RTS. This is given
at the beginning of section III where the formula is then specialized
to the case of Mittag-Leffler dichotomous noise. A comparison between
analytical results and Monte Carlo simulations concludes this section.
Section IV contains a summary of results as well as some directions
for future work. 

\begin{figure}[h]
\includegraphics[width=15 cm]{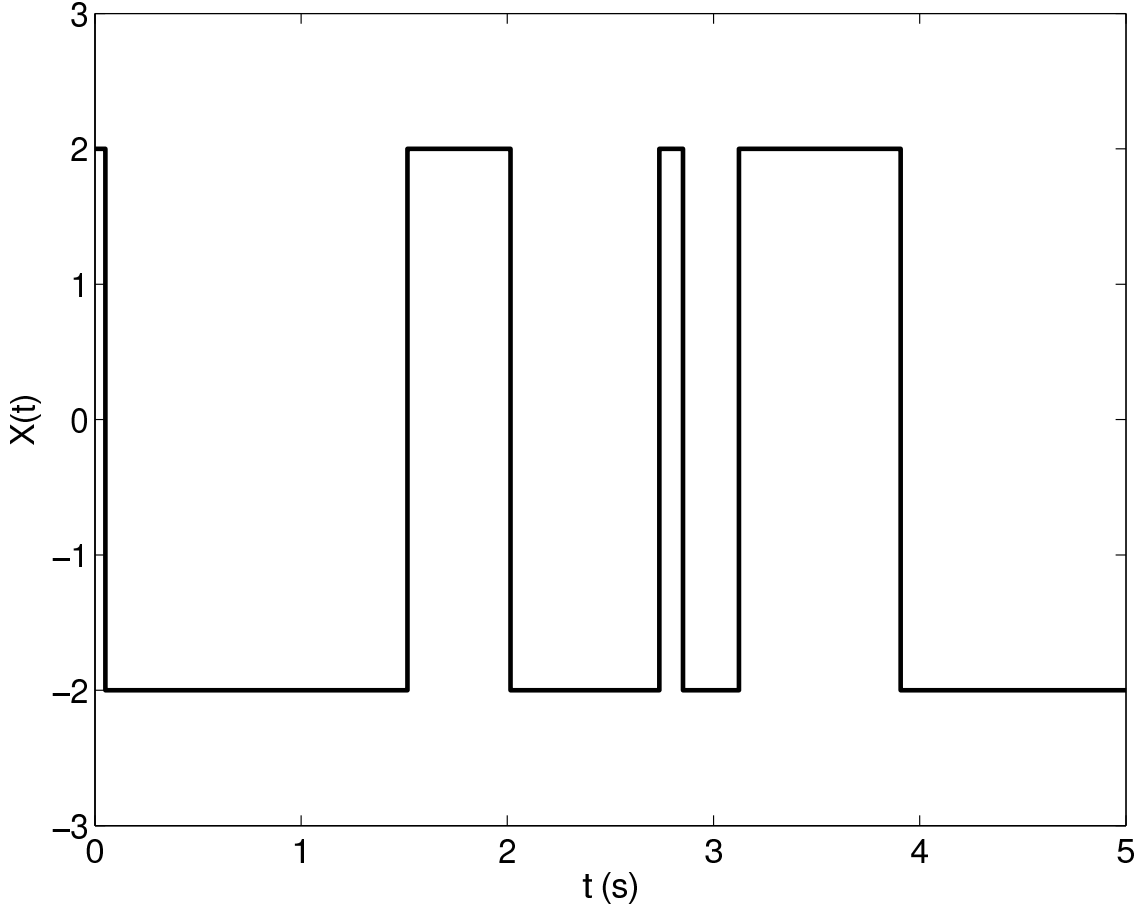}% Here is how to import EPS art
\caption{\label{fig:0} Graphical representation of a RTS.}
\end{figure}

\section{Theory}

\subsection{The stationary random telegraph signal}

The {\em random telegraph signal} is a simple {\em compound renewal process}. A random variable
$X(t)$ can assume two opposite values, say $X(t) = \pm a$, it starts with value, say, $X(t) = +a$ for
$t=t_0=0$ and it suddenly changes value assuming the opposite one at time instants $t_1,\ldots,t_n,\ldots$
with the differences $\tau_i = t_i - t_{i-1}$ independently and exponentially distributed with common intensity $\mu$.
The probability density function of these differences is $\psi(\tau) = \mu \exp(-\mu \tau)$.
If $N(t)$ denotes the number of switches up to time $t$, the probability $P(N(t)=n)$ of having
$n$ jumps up to time $t$ is given by the Poisson distribution: $P(n,t)=P(N(t)=n)=\exp(-\mu t) (\mu t)^n/n!$.
The random telegraph signal is a time-homogeneous stochastic process, therefore the
autocovariance function:
\begin{equation}
\label{autocorr}
C_{XX} (\Delta t;t) = \mathbb{E} [X(t) X(t+\Delta t)]
\end{equation}
does not depend on $t$, but only on the lag $\Delta t$.

\subsection{Essentials of renewal theory}

A way to generalize exponential waiting times and the Poisson process is provided by
{\em renewal theory}.
A {\em renewal process} is a particular instance of one-dimensional {\em point process}. 
Renewal events occur at consecutive times $t_0=0, t_1, \ldots, t_n, \ldots$ and the differences
$\tau_i = t_i - t_{i-1}$ are independent and identically distributed random variables.
They are called {\em durations} or {\em waiting times}.
The {\em counting process} $N(t)$ related to a renewal process is the positive integer
giving the number of renewals up to time $t$. $P(n,t)$ denotes the probability of
observing $N(t) = n$. Renewal theory is discussed in great detail in the book by Feller
\cite{feller} and in two monographs, one by Cox and the other by Cox and Isham \cite{cox}. 
An account for physicists also exists \cite{godreche}.

The probability distribution $P(n,t)$ can be related to the probability density, $\psi(\tau)$, of waiting times. 
It is sufficient to observe that $t_n$, the instant of the $n$-th renewal event
is also a random walk (in this case the sum of i.i.d. positive random variables):
\begin{equation}
\label{renewal}
t_n = \sum_{i=1}^{n} \tau_i,
\end{equation}
and the probability of having $n$ jumps up to time $t$ coincides with the
probability of $t_n \leq t$ ($n$ jumps up to $t_n$ and no jumps from $t_n$ to $t$):
\begin{equation}
\label{distribution1}
P(n,t) = [\Psi * \psi^{*n}] (t)\,,
\end{equation}
where $*$ represents the convolution operator, $\psi^{*n}(\tau)$ is the $n$-fold convolution of
the density $\psi(\tau)$ and $\Psi(\tau) = 1 - \int_{0}^{\tau} \psi(u)\,du$ is the
complementary cumulative distribution function (a.k.a. survival function).
A continuous time random walk (CTRW) with exponentially-distributed waiting times and Poisson-distributed
counting process is Markovian, moreover it has stationary and independent increments and,
therefore, belongs to the class of L\'evy processes \cite{feller,hoel,scalas06}. The exponential distribution
is the only continuous memoryless distribution. In this case, the counting probability distribution
$P(n,\Delta t; t)$ of observing $n$ jumps from a generic instant $t$ to instant $t + \Delta t$ 
coincides with $P(n,\Delta t)$, that is the probability of observing $n$ jumps 
from $t=0$ to $\Delta t$. In the general case, this is not true and $P(n,\Delta t; t)$ explicitly
depends on $t$ \cite{cox}. In order to obtain $P(n,\Delta t; t)$, one needs to know the distribution
of the {\em forward recurrence time} a.k.a 
{\em residual life-time} $y$, the time interval between $t$ and the next renewal event. If $g(y;t)$
denotes the probability density function of the residual life-time, one has
\begin{equation}
\label{distribution2}
P(n,\Delta t;t) = [g * \Psi * \psi^{*(n-1)}] (\Delta t; t) = \int_{0}^{\Delta t} g(y;t) P(n-1,\Delta t-y)\,dy\,,
\end{equation} 
because the probability of $n$ jumps occurring from instant $t$ to instant $t+\Delta t$ is 
given by the probability of $t + t_{(n-1)} \leq t+\Delta t$ and $t+t_{(n-1)} = 
t+\sum_{i=1}^{n-1} \tau_i$.
The probability density function of the residual life-time $g(y;t)$ can be written in terms
of the {\em renewal density} $h(t)$. The renewal density is the time derivative of the average
number of counts up to time $t$, the so-called {\em renewal function}, $H(t)$,
that is the average value of the random variable $N(t)$: $H(t) := \mathbb{E}[N(t)]$, and
$h(t) := d H(t)/ d t$. The renewal density is a measure of the activity
of the renewal process. For the Poisson process and exponentially distributed
waiting times $h(t)$ is a constant and coincides with the inverse of the
average waiting time or the average number of events per time unit, in
other words, $h(t) = \mu$. It can be shown that
\cite{cox}
\begin{equation}
\label{renewalequation}
h(t) = \psi(t) + \int_{0}^{t} h(t-u) \psi(u) \, du\,,
\end{equation}
and that the density
$g(y;t)$ is given by \cite{cox}
\begin{equation}
\label{lifetimedistribution}
g(y;t) = \psi(t+y) + \int_{0}^{t} h(t-u) \psi (u+y) \, dy.
\end{equation}

\subsection{A renewal process of Mittag-Leffler type}

F. Mainardi and R. Gorenflo, together with one 
of the authors of the present paper have studied the renewal process of Mittag-Leffler type \cite{sgm}.
It is characterized by the following survival function
\begin{equation}
\label{ml}
\Psi (\tau) = E_{\beta} (-\tau^{\beta}),
\end{equation}
where $E_{\beta} (z)$ is the one-parameter Mittag-Leffler function
\begin{equation}
\label{opml}
E_{\beta} (z) = \sum_{n=0}^{\infty} \frac{z^n}{\Gamma(\beta n +1)}
\end{equation}
where $\Gamma(x)$ is Euler gamma function.
The Mittag-Leffler function is a legitimate survival function for $z = -\tau^{\beta}$,
$\tau \geq 0$,
$0< \beta \leq 1$ and coincides with the exponential for
$\beta = 1$. The Mittag-Leffler survival function interpolates
between a stretched exponential for small values of $\tau$ and
a power law for large values of $\tau$. In particular, for $0\leq \tau << 1$ one has:
\begin{equation}
\label{smalltau}
E_{\beta} (-\tau^{\beta}) \simeq 1 - \frac{\tau^{\beta}}{\Gamma(\beta +1)} \simeq 
\exp(-\tau^{\beta}/\Gamma(\beta+1)),
\end{equation}
and for $\tau \to \infty$
\begin{equation}
\label{largetau}
E_{\beta} (-\tau^{\beta}) \simeq \frac{\sin(\beta \pi)}{\pi} \frac{\Gamma(\beta)}{\tau^{\beta}}.
\end{equation}

\section{Results}

\subsection{Autocovariance for a general RTS}

Let us now consider a generalized random telegraph signal 
with durations that do not necessarily follow the
exponential distribution. As the signal has zero mean,
the autocovariance function $C_{XX} (t) = \mathbb{E}[X(t) X(t+\Delta t)]$ coincides
with the autocorrelation function. Throughout this paper, autocovariance functions
will not be normalized.
The generalized RTS is not a time-homogeneous
(stationary) random process and the autocovariance
function does depend on the initial time of evaluation. In particular,
one gets
\begin{equation}
\label{autocorrns}
C_{XX} (\Delta t; t) = \mathbb{E} [X(t) X(t+\Delta t)]= a^2 \sum_{n=0}^{\infty} (-1)^n P(n,\Delta t; t).
\end{equation}
Equation (\ref{autocorrns}) is a consequence of the fact that, in the time
interval between $t$ and $t+\Delta t$, with probability $P(n,\Delta t; t)$,
there can be an arbitrary, but finite, integer number $n$ of renewal events where the
process of amplitude $a$ changes sign. As $P(n,\Delta t; t)$ is available from
equations (\ref{distribution2}), (\ref{renewalequation}) and (\ref{lifetimedistribution}), in principle, one can compute the {\em time
dependent} autocovariance from the knowledge of the distribution
of waiting times. However, the use of convolutions is painful and some analytical
progress is possible by means of the method of Laplace transforms.
Given a sufficiently well-behaved (generalized) function $f(t)$ with
non-negative support, the Laplace transform is given by
\begin{equation}
\label{laplace}
\tilde{f} (s) = \int_{0}^{\infty} \exp(-st) f(t) \, dt.
\end{equation}
(A generalized function is a distribution in the sense of Sobolev and
Schwartz \cite{gelfand}).

Here, the method of the double Laplace transform described by Cox turns
out to be very useful \cite{cox}. Let us denote by $s$ the variable
of the Laplace transform with respect to $\Delta t$ or $y$ and by $u$
the variable of the Laplace transform with respect to $t$. Then
$\tilde{\tilde{g}}(s;u)$, the double Laplace transform of the
residual life-time density $g(y;t)$ in (\ref{lifetimedistribution}),
is given by
\begin{equation}
\label{doublelaplaceresidual}
\tilde{\tilde{g}} (s,u) = \frac{(\tilde{\psi} (s) - \tilde{\psi} (u))(1+\tilde{h} (u))}{u-s}.
\end{equation}
Now, one gets from equation (\ref{distribution2}) that
\begin{equation}
\label{doublelaplacedistribution}
\tilde{\tilde{P}} (n, s; u) = \tilde{\tilde{g}} (s;u) [\tilde{\psi}(s)]^{n-1} \tilde{\Psi} (s)
= \frac{(\tilde{\psi} (s) - \tilde{\psi} (u))(1+\tilde{h} (u))}{u-s} [\tilde{\psi}(s)]^{n-1} \tilde{\Psi} (s).
\end{equation}
The double inversion of the Laplace transforms in equation (\ref{doublelaplacedistribution}) may be
as a formidable task as the direct calculation of convolutions. However, equation 
(\ref{doublelaplacedistribution}) can be used to investigate the asymptotic
behaviour of $P(n,\Delta t;t)$ for small and large values of $t$ by means
of Tauberian-like theorems \cite{feller}. If the average waiting
time $\langle \tau \rangle = \mathbb{E}[\tau]$ is finite, then
for $t \to \infty$, as a consequence of the
{\em renewal theorem}, one has that $h(t) \to 1/\langle \tau \rangle$ and the renewal
process behaves as a Poisson process \cite{cox}. This is not the case for the
renewal process of Mittag-Leffler type where $h(t) = t^{\beta -1}/\Gamma(\beta)$
(and $H(t) = t^{\beta}/\Gamma(1+\beta)$). In other words, all the RTSs following
a renewal process with $\mathbb{E}[\tau]<\infty$ reach a {\em stationary state}
after an initial non-stationary transient; in this stationary state, the autocovariance
does no longer depend on $t$. The RTS of Mittag-Leffler
type never reaches such a state.

\subsection{Analytical result for the RTS of Mittag-Leffler type}

Let us consider the case $t=0$. Equation (\ref{autocorrns})
simplifies to
\begin{equation}
\label{autocorrml0}
C_{XX} (\Delta t; 0|\beta)=a^2 \sum_{n=0}^{\infty} (-1)^n P(n,\Delta t).
\end{equation}
Using the Laplace transform of the Mittag-Leffler survival function (\ref{ml}) $s^{\beta-1}/(1+s^{\beta})$
and of the corresponding density function $1/(1+s^{\beta})$ \cite{sgm,podlubny}, one gets:
\begin{equation}
\label{autocorrml0lt}
\tilde{C}_{XX} (s;0|\beta) = a^2 \frac{s^{\beta-1}}{2+s^{\beta}};
\end{equation}
this Laplace transform can be inverted to yield
\begin{equation}
\label{autocorrml0an}
C_{XX} (\Delta t;0|\beta) = a^2 E_{\beta} (-2 (\Delta t)^{\beta}).
\end{equation}
For $\beta =1$ this formula reduces to
\begin{equation}
\label{autocorrRTS}
C_{XX} (\Delta t;0|\beta=1) = a^2 \exp(-2 \Delta t).
\end{equation}
Indeed, equation (\ref{autocorrRTS})
can be directly derived for a RTS with intensity $\mu=1$. In this case,
$P(n,\Delta t) = \exp(-\Delta t) (\Delta t)^{n}/n!$, therefore from equation (\ref{autocorrml0}) one gets
\begin{equation}
\label{autocorrRTS1}
C_{XX} (\Delta t;0|\beta=1) = a^2 \exp(-\Delta t) \sum_{n=0}^{\infty} (-1)^n \frac{(\Delta t)^n}{n!} = a^2 \exp( - 2 \Delta t).
\end{equation}
Thus, equation (\ref{autocorrml0an}) coincides with the autocovariance function for the usual 
RTS when $\beta = 1$.

\subsection{Monte Carlo simulations}

The essential ingredient for Monte Carlo simulation of a RTS of Mittag-Leffler type is the generation
of Mittag-Leffler deviates. An efficient transformation
method has been thoroughly discussed in \cite{fulger08} based on
the theory of geometric stable distributions \cite{pakes,kozubowski,kotz}.
Mittag-Leffler distributed random numbers can be obtained using
the following formula \cite{rachev}:
\begin{equation}
\label{rachev}
\tau_{\beta} = -\ln u \left( \frac{\sin(\beta \pi)}{\tan(\beta \pi w)} - \cos(\beta \pi) \right)^{\frac{1}{\beta}}
\end{equation}
where $u$ and $w$ are real random numbers uniformly distributed
in $(0,1)$. For $\beta =1$ eq. (\ref{rachev}) gives exponentially
distributed waiting times with $\mu = 1$.

Once a method for generating Mittag-Leffler deviates is available, 
the Monte Carlo simulation of an uncoupled CTRWs is  
straightforward. It is sufficient to generate a sequence of $n(t) + 1$ independent and identically distributed waiting 
times $\tau_{\beta,i}$ until their sum is greater than $t$. Then the last waiting time can be
discarded and $n(t)$ alternate signs are generated. 
A Monte Carlo realisation starts, say, at $X(0) = +a$, then
at time $t_1 = \tau_{\beta,1}$ the process flips to the new
value $X(t_1) = -a$ and stays there for $t_1 < t <t_2$ with
$t_2 = \tau_{\beta,1} + \tau_{\beta,2}$, and so on.

In Fig. 2, Monte Carlo estimates of the autocovariance function $C_{XX} (\Delta t;0|\beta)$
are compared to formula (\ref{autocorrml0an}) for several values of $\beta$
and $a=2$. The estimates have been produced using an {\em ensemble}
average estimator. For each of 10000 values of $\Delta t$, $1000$ values of the product
$X(0)X(\Delta t)$ have been produced. The sampling frequency is assumed to be 1/1000, so
that the total length of the signal is 10 arbitrary units (seconds). A direct eye inspection
shows that the agreement between simulations and theory is excellent.

Fig. 3 shows the effect of non-stationarity. The autocovariance
functions $C_{XX} (\Delta t;t|\beta)$ are compared for 
$\beta = 0.8$, $a=2$ and several values of $t$. For $t=0$,
the theoretical line is plotted. The dependence of the autocovariance
function on the choice of $t$ points to the fact that standard 
spectral methods routinely used in signal analysis
would fail in this case.

\begin{figure}[h]
\includegraphics{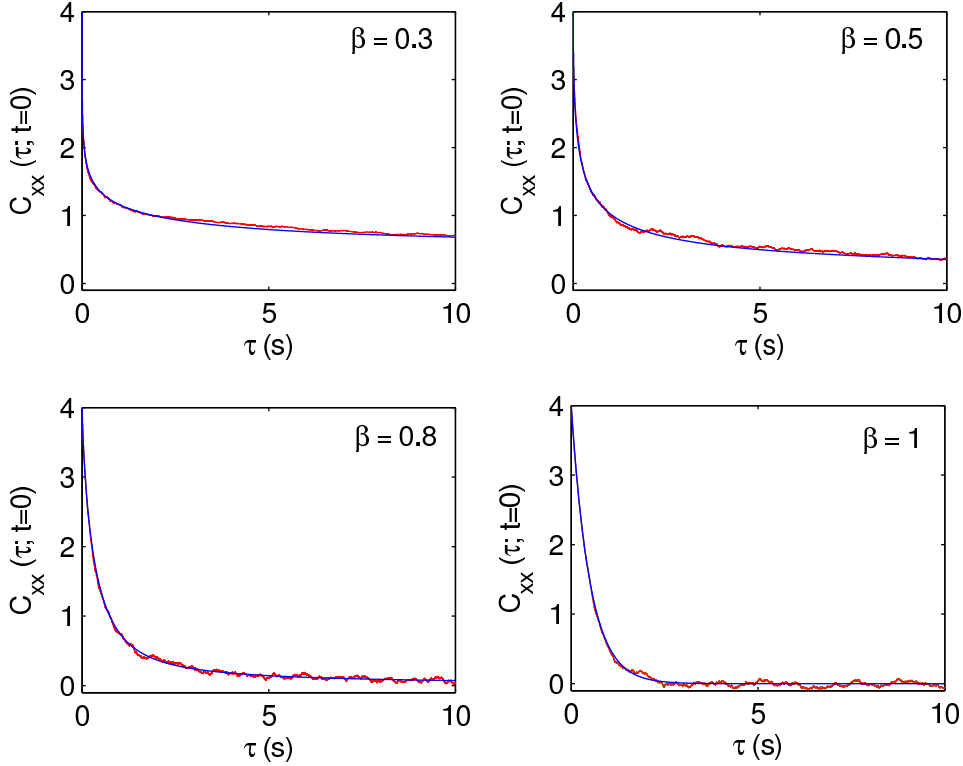}% Here is how to import EPS art
\caption{\label{fig:1a} (Color online) Monte Carlo estimates of the autocovariance function $C_{XX} (\Delta t;0|\beta)$
(red irregular lines) compared to the exact result in equation (\ref{autocorrml0an}) (blue smooth
lines) for several values of $\beta$
and $a=2$. Time is in arbitrary units called seconds. 
The autocovariance functions are not normalized.}
\end{figure}

\begin{figure}[h]
\includegraphics{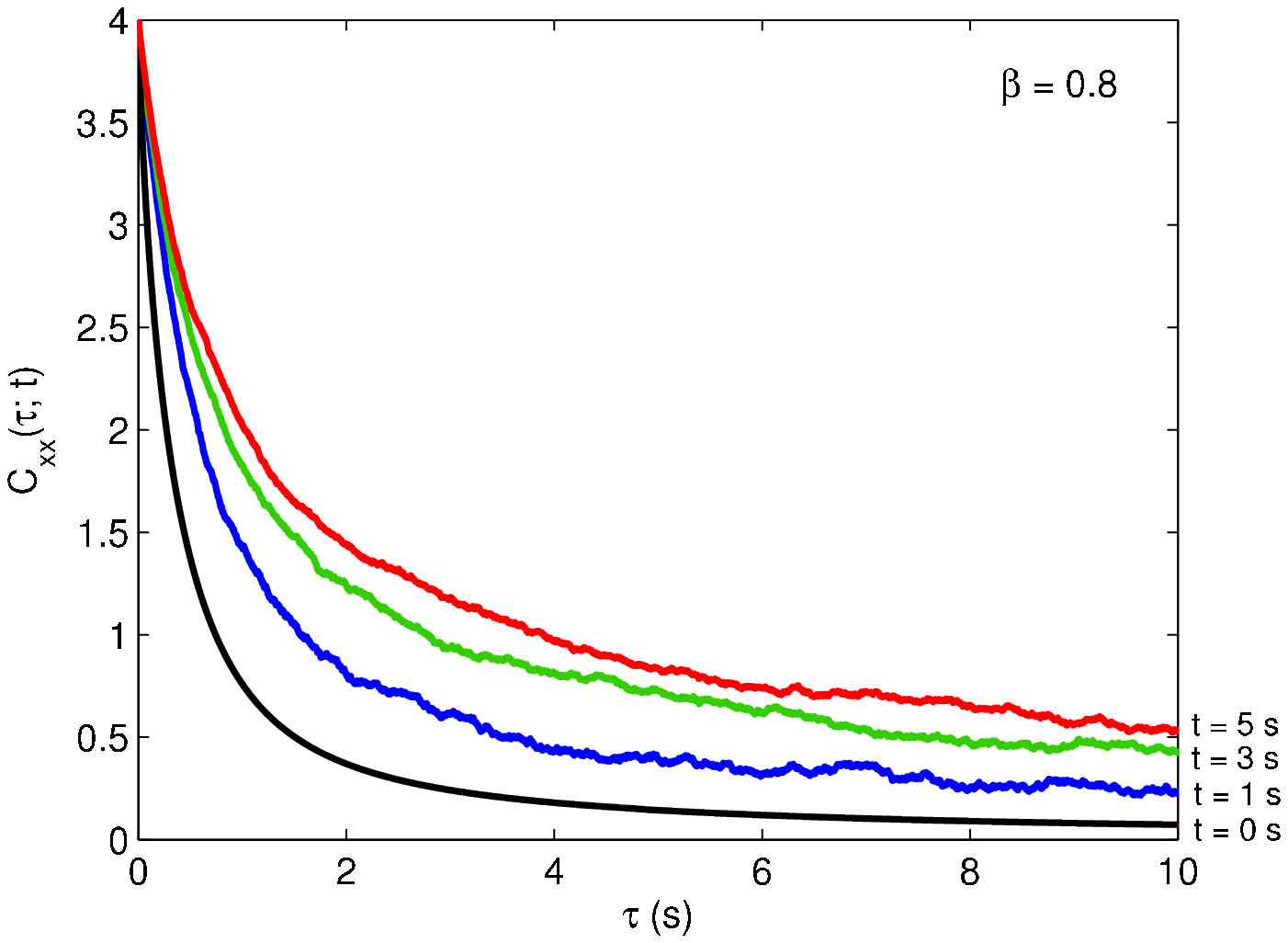}% Here is how to import EPS art
\caption{\label{fig:1b} (Color online) Non-normalized autocovariance
functions $C_{XX} (\Delta t;t|\beta)$ for 
$\beta = 0.8$, $a=2$ and several values of $t$. Time is measured in
arbitrary units called seconds.}
\end{figure}

\section{Discussion and conclusions}

%\subsection{Summary}

The main result of this paper is given in equation (\ref{autocorrns}) for
the autocovariance function of a generalized RTS, the double Laplace
transform of the probability distribution $P(n,\Delta t;t)$ 
is given in (\ref{doublelaplacedistribution}); these results are further
specialized to a RTS of Mittag-Leffler type in equation (\ref{autocorrml0an}). The derivation
of such equations is based on renewal theory.
The analytical result in equation (\ref{autocorrns}) is then successfully compared to Monte Carlo
simulations.

%\subsection{Discussion and outlook}

An important feature of a RTS of Mittag-Leffler type is that the renewal
density $h(t)$ of the Mittag-Leffler renewal process vanishes for 
$t \to \infty$ for $0 < \beta < 1$. Therefore, the hypoteses of the renewal theorem are
not satisfied and the stationary regime is never reached. Therefore,
contrary to other renewal processes, such as the Weibull process,
that, after a transient, become stationary, the renewal process
of Mittag-Leffler type is non-stationary also if it is observed
far from its beginning. In particular, standard spectral methods
fail to describe the features of the RTS of Mittag-Leffler type.

These considerations may be useful when studying the origin of 
1/$f$ noise, a problem not addressed here but which will be
the subject of a future paper. Indeed, the renewal process of Mittag-Leffler
type can be described as an infinite mixture of exponential distributions,
for instance, one has for the survival function
\begin{equation}
\label{mixture1}
E_{\beta} \left( -t^\beta \right) = \int_{0}^{\infty} g(\mu) \exp(-\mu t) d \mu,
\end{equation}
where
\begin{equation}
\label{mixture2}
g(\mu) = \frac{1}{\pi} \frac{\sin(\beta \pi)}{\mu^{1+\beta}+2 \cos(\beta \mu)\mu + \mu^{1-\beta}}.
\end{equation}
Therefore, based on the ideas developed in \cite{mcwhorter55,stepanescu74,milotti}, one already expects to
have a power spectrum following a power law. Using the fact that
\begin{equation}
\label{power}
\hat{S}_{XX} (\omega;0|\beta) = {\cal F} [C_{XX} (\Delta t;0|\beta)] = \int_{-\infty}^{+\infty} E_{\beta}
\left( -2 |v|^{\beta} \right) \exp(i \omega v) dv,
\end{equation}
one gets
\begin{equation}
\label{powerlaplace}
\hat{S}_{XX} (\omega;0|\beta) = 2 \mathrm{Re} [ \tilde{C}_{XX} (s = - i \omega;0|\beta)],
\end{equation}
and the power spectrum for $t=0$ can be evaluated from equation (\ref{autocorrml0lt}). 
It turns out that
\begin{equation}
\label{psd0}
\hat{S}_{XX} (\omega;0|\beta) = 2 a^2 \mathrm{Re} \left[ 
\frac{(-i \omega)^{\beta -1}}{2+(-i \omega)^\beta}  \right] =
a^2 \frac{4 \omega^{\beta -1} \sin (\beta \pi/2)}{4+4 \omega^{\beta} \cos
(\beta \pi/2) + \omega^{2 \beta}}
\end{equation}
Once
again, it must be stressed that, due to non-stationarity, the power spectrum depends on $t$ and not only on the lag
$\Delta t$.

\begin{acknowledgments}
This paper has been supported by local research grants for the years 2007 and 2008. E.S. acknowledges
support of PRIN 2006 Italian funds.
\end{acknowledgments}


\begin{thebibliography}{99}
\bibitem{kolmogorovnote}
Kolmogorov's theorem is the basis for existence (and uniqueness) theorems of stochastic processes. It is
discussed in sections 3 and 4 of reference \cite{kolmogorov33}.
\end{thebibliography}
\end{document}